**Reinforcement Learning based Cyberattack Model for Adaptive Traffic Signal Controller in Connected Transportation Systems**

**Muhammad Sami Irfan***
Ph.D. Student
Department of Civil, Construction & Environmental Engineering
The University of Alabama
3013 Cyber Hall, Box 870205, 248 Kirkbride Lane, Tuscaloosa, AL 35487
Tel: (205) 239-9705; Email: mirfan@crimson.ua.edu

**Mizanur Rahman, Ph.D.**
Assistant Professor
Department of Civil, Construction & Environmental Engineering
The University of Alabama
3015 Cyber Hall, Box 870205, 248 Kirkbride Lane, Tuscaloosa, AL 35487
Tel: (205) 348-1717; Email: mizan.rahman@ua.edu

**Travis Atkison, Ph.D.**
Associate Professor
Department of Computer Science
The University of Alabama
3057 Cyber Hall, Box 870290, Tuscaloosa, AL 35487
Tel: (205) 348-4740; Email: atkison@cs.ua.edu

**Sagar Dasgupta**
Ph.D. Student
Department of Civil, Construction & Environmental Engineering
The University of Alabama
3013 Cyber Hall, Box 870205, 248 Kirkbride Lane, Tuscaloosa, AL 35487
Tel: (864) 624-6210; Email: sdasgupta@crimson.ua.edu

**Alexander Hainen, Ph.D.**
Associate Professor
Department of Civil, Construction & Environmental Engineering
The University of Alabama
3010 Cyber Hall, Box 870205, 248 Kirkbride Lane, Tuscaloosa, AL 35487
Tel: +1 231 883 2669; Email: ahainen@eng.ua.edu

*Corresponding author

Word count:  6,228 words text + 2 table x 250 words (each) = 6,728 words

Submission date: August 1, 2022

Paper submitted for presentation at the Transportation Research Board 102nd Annual Meeting and for publication in Transportation Research Record

## ABSTRACT

In a connected transportation system, adaptive traffic signal controllers (ATSC) utilize real-time vehicle trajectory data received from vehicles through wireless connectivity (i.e., connected vehicles) to regulate green time. However, this wirelessly connected ATSC increases cyber-attack surfaces and increases their vulnerability to various cyber-attack modes, which can be leveraged to induce significant congestion in a roadway network. An attacker may receive financial benefits to create such a congestion for a specific roadway. One such mode is a 'sybil' attack in which an attacker creates fake vehicles in the network by generating fake Basic Safety Messages (BSMs) imitating actual connected vehicles following roadway traffic rules. The ultimate goal of an attacker will be to block a route(s) by generating fake or 'sybil' vehicles at a rate such that the signal timing and phasing changes occur without flagging any abrupt change in number of vehicles. Because of the highly non-linear and unpredictable nature of vehicle arrival rates and the ATSC algorithm, it is difficult to find an optimal rate of sybil vehicles, which will be injected from different approaches of an intersection. Thus, it is necessary to develop an intelligent cyber-attack model to prove the existence of such attacks. In this study, a reinforcement learning based cyber-attack model is developed for a waiting time-based ATSC. Specifically, an RL agent is trained to learn an optimal rate of sybil vehicle injection to create congestion for an approach(s). Our analyses revealed that the RL agent can learn an optimal policy for creating an intelligent attack.

## INTRODUCTION

Traffic signal controllers (TSC) are an integral part of transportation systems and are responsible for efficient signal timing to control movements of traffic flow by avoiding conflicts. Traditional fixed time TSCs switch between phases on fixed-time interval basis. They are pre-programmed and the phase durations remain unchanged regardless of the volume of traffic. This relative inflexibility of fixed time signal has necessitated the introduction of actuated TSCs. The actuated TSC utilizes sensors to detect vehicle arrival rates. Sensors employed for this purpose include inductive loop detectors, magnetometers, magnetic detectors, microwave radars, infrared detectors, ultrasonic detectors and video image processors [1]. Depending on the volume of traffic sensed by these sensors, the actuated TSC can change its cycle time to accommodate the greater traffic volume accordingly. With the advent of new technologies, a new paradigm in TSCs have emerged in the form of adaptive traffic signal controllers (ATSC).

In contrast with other types of TSCs, ATSCs are able to dynamically change their signal plans in response to traffic demand. This can be done through the implementation of various algorithms within the TSC that can find the optimum signal plan corresponding to the arrival of vehicles and waiting time at the intersection. Recently, CV technology has facilitated the development of ATSCs as CVs are able to wirelessly communicate to the roadside infrastructure and deliver their trajectory information in the form of Basic Safety Messages (BSMs). The ATSCs are able to receive the broadcasted trajectory information from the vehicles and optimize their signal plans accordingly using that data. However, this wirelessly connected ATSC increases cyber-attack surfaces and increases their vulnerability to various cyber-attack modes, which can be leveraged to induce significant congestion in a roadway network.

An attacker may receive financial (or political) benefits to create such a congestion for a specific roadway. One such mode is a 'sybil' attack in which an attacker creates fake vehicles in a roadway network by broadcasting fake Basic Safety Messages (BSMs) following Society of Automotive Engineers (SAE) J2735 BSM standard as well as IEEE 1609.3 standard, and imitating actual connected vehicles following roadway traffic rules. Note that an attacker can acquire an on-board unit because of availability in the market and can broadcast BSMs following SAE J2735 and IEEE 1609.3 standards. The ultimate goal of an attacker will be to block a route(s) by generating fake or 'sybil' vehicles at a rate such that the signal time and phasing changes occur without flagging any abrupt change in number of vehicles. Because of the highly non-linear and unpredictable nature of vehicle arrival rates and the ATSC algorithm, it is difficult to find an optimal rate of sybil vehicles, which will be injected from different approaches of an intersection. Thus, it is necessary to develop an intelligent cyber-attack model to prove the existence of such attacks so that a robust ATSC can be developed considering their vulnerabilities.

Several researchers have investigated the vulnerabilities within CV based ATSCs in depth [2,6,7,8,9,10]. Falsified data attacks or data spoofing has been particularly effective against ATSC systems. It was found that delay-based ATSC schemes were more vulnerable compared to other schemes [9]. A prior work had implemented an intelligent Reinforcement Learning based attack against an actuated TSC system in [10]; however, such an attack against an ATSC is yet to be found in the literature.

In this study, a reinforcement learning based cyber-attack model is developed for a waiting time-based ATSC. Specifically, an RL agent is trained using vehicle trajectory data from different approaches of a signalized intersection to determine an optimal rate of sybil vehicle injection to create congestion for an approach(s). By injecting fake or "sybil" vehicles into the regular traffic flow, we aim to fool the ATSC and thereby create congestion at an intersection. Furthermore, our developed attack model utilizes an intelligent, automated attack agent which can learn to create congestion at the subject intersection without requiring supervision from a human attack agent. The contributions of this paper are as follows: (i) develop a cyber-attack model for waiting time based ATSC using deep RL for the first time; (ii) defined a new reward function for the RL so that an optimal rate of sybil vehicle can be injected to create congestion for an approach(s) of a signalized intersection; (ii) develop a fake vehicle removal strategy to simulate sybil attack realistically in a simulation environment for the first time.

## RELATED WORKS

Several authors have investigated the issue of vulnerabilities in TSC systems. TSC systems can be physically compromised by the attacker to gain access and thereby tamper with the system. However, this involves the attacker to be physically present in the vicinity of the traffic cabinet and equipment. This poses immediate risk of detection, so it is not feasible for the attacker [5]. On the other hand, an attack can also be carried out wirelessly. The authors in [5] investigated a wireless attack scheme on an intelligent wireless traffic management system. They were able to use a 5.8GHz radio to wirelessly connect to a traffic signal controller from a remote location. Upon connecting to the network, they could access the controller via open debug ports and through remote control capabilities of the controller. The access to the controllers enabled them to reduce or elongate the duration of a particular light state in order to achieve their desired attack objectives.

In the connected vehicle (CV) environment, the vehicles themselves transmit their trajectory data wirelessly in the form of BSMs via their On-Board Units (OBUs). The CV based ATSC systems utilize the received trajectory information to actuate the signals and optimized traffic signal parameters. An attacker can utilize a compromised OBU of a CV and use it to transmit falsified trajectory information to the TSC. As a result of these falsified data transmissions, the TSC can be fooled in such a way as to produce suboptimal signal timing plan [6].

Based on such vulnerabilities, the authors in [7] developed an attack scenario which they named as "black-box" attack. The attack treats the ATSC as a black box and the attacker is assumed to have no access to the control logic. Instead, a surrogate model is trained to learn the hidden control logic within a TSC. From this model, critical traffic features are identified which are used by the TSC to optimize the signal timing plans. Knowledge of these features can enable an attacker manipulate traffic in such a way which can cause the TSC to produce unoptimized signal timings. A similar black box approach was used in [8] where they showed that group of colluding vehicles can send falsified data to the TSC in order to reduce their waiting time. The demonstrated attack was against a Deep Reinforcement Learning (DRL) based ATSC and the adversary was DRL based as well. Other authors have also explored DRL based attack agents that can independently learn to achieve attack objectives. A DRL based agent was employed in [10] to create an attack on an actuated TSC. Such an agent was found to be highly effective in increasing the total waiting time at the intersection. The advantage of this attack is that it can be carried out remotely and with minimal intervention required from the attacker's perspective. However, the model used in the attack had several limitations. Firstly, the TSC systems against which the attacks were deployed were not ATSC systems but rather actuated ones. Secondly, while their model involved injecting fake or "sybil" vehicles into the traffic network, it had no arrangements to prevent the interference between the real vehicles and the fake vehicles. Since the fake vehicles are essentially falsified BSM messages sent wirelessly, they have no physical presence on the road and should in no way interfere or interact with the physical vehicles in their travel path. Also, an in-depth analysis of the attack impact at the individual approaches of the intersection was not presented.

Against these existing attacks against TSC, an impact evaluation of falsified data attacks was performed on several backpressure-based TSC algorithms. It was found that delay-based schemes were particularly susceptible to time spoofing attacks [9]. Based on this finding, a "slow poisoning" attack was demonstrated in [2] against a waiting time based ATSC. This involved an attacker slowly injecting fake or "sybil" vehicles into the traffic stream at a particular approach in order to corrupt the signal timing of the ATSC. However, this approach involved a trial-and-error method for identifying the optimum rate of vehicle injection. As such, it could not be applied to several approaches at the same time.

Thus, the pre-existing body of literature has several demonstrations of attacks against TSC systems. However, to the best of the authors knowledge there has been no automated attacks against waiting time based ATSC systems that can create congestion at the intersection level. This study presents a novel DRL based attack against a waiting time based ATSC.

## ATTACK MODELING

The developed attack model relies on random sybil vehicle injection at the various approaches in order to create the congestion. Since the attacker also wants to avoid immediate detection by the ATSC algorithm, the attack needs to be done gradually at a slow rate. Such an attack was termed as a "slow-poisoning" attack in literature [2]. This results in an attack which requires extensive number of hours to propagate. Performing this attack manually would be cumbersome and would require a lot of time on the part of the attacker. Thus, the motivation of implementing an RL agent to carry out a sybil attack against the proposed waiting time based adaptive traffic signal controllers is two-folds:

- To automate the process of carrying out the attack without requiring extended periods of supervision by a human attacker.
- To learn the optimal sybil vehicle injection rate over time to maximize congestion at the intersection.

In reinforcement learning, an agent interacts with an environment and observes its state. The agent's interaction with the environment results in a change in its state. After each interaction and state observation, the agent receives a reward based on the effectiveness of the interaction. If the change in state as a result of the interaction is conducive to the agent's goal, a reward is assigned to the agent. On the other hand, if an unfavorable change occurs, the agent will be assigned no reward or even a negative reward. Over time, the agent aims to maximize its cumulative reward received. In this way, the RL agent identifies the correct actions to take based on any observed environment state.

In the context of our attack problem, the intersection is the environment within which the RL agent interacts. The RL agent interacts with the environment by taking various actions which correspond to injecting vehicles at the different approaches of the intersection. The different components of the RL system are modeled as follows for the current problem:

### Environment

The environment is the world within which the RL agent interacts through executing actions and retrieving state information. For our developed attack model, the agent is concerned with attacking the ATSC at one particular intersection, which we refer to as the subject intersection. The agent reads various state information of the ATSC. It also takes actions based on the current state which comprise of inserting fake vehicles at the four approaches connected to the subject intersection. The four approaches are in-turn sub divided into through and left turn movements resulting in the following eight traffic movements- east bound through (EBT), east bound left (EBL), west bound through (WBT), west bound left (WBL), south bound through (SBT), south bound left (SBL), north bound through (NBT), and north bound left (NBL). Thus, the environment of our agent consists of the subject intersection and the four approaches connected to it.

### State space

The state space, S represents the set of variables that the agent uses to perceive its environment, in this case being the subject intersection. For the attack model, we consider the number of vehicles at each of the four approaches of the intersection, the current active phase, and the remaining duration of the current phase. These variables are combined into one array which form the state space for the RL problem.

**Action space**

The action space for our RL formulation represents the possible set of actions that the agent can take to interact with the environment. In our RL problem, the environment is defined as a four-way intersection as shown in **Figure 1**. Each approach of the intersection has three lanes: a dedicated left-turn lane, a dedicated through lane and shared through and right turn lane. Our agent interacts with this environment by injecting sybil vehicles at the various approaches in each step of the simulation. The yellow arrows depict the injection point of fake vehicles into the traffic network by the attacker. It is assumed that due to resource constraints, the attacker is only able to inject a maximum of two vehicles at any given time. Thus, the possible actions our agent can take are represented in **Table 1** below.

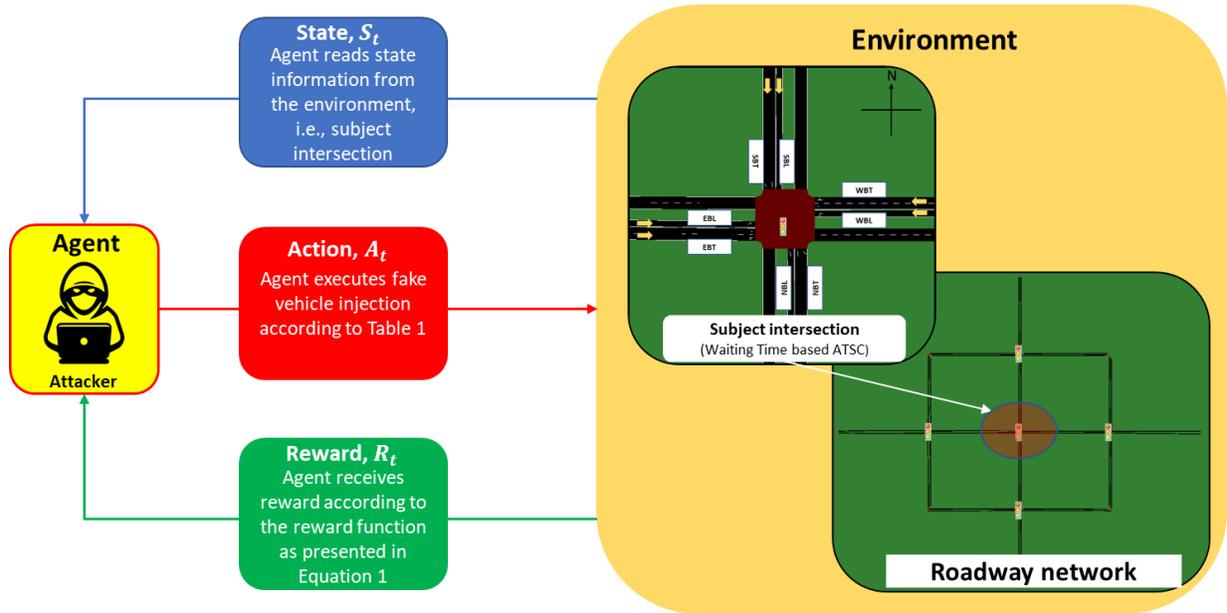

**Figure 1 Subject intersection showing the individual traffic movements**

**Table 1 Action modes and corresponding executable actions by the agent**

| Action mode | Executable |
|---|---|
| **0** | No vehicles injected |
| **1** | 1 vehicle injected at start of West bound through approach |
| **2** | 1 vehicle injected at start of East bound through approach |
| **3** | 1 vehicle injected at start of South bound through approach |
| **4** | 1 vehicle injected at start of North bound through approach |
| **5** | 1 vehicle injected at start of West bound through approach **AND** 1 vehicle injected at start of East bound through approach |
| **6** | 1 vehicle injected at start of South bound through approach **AND** 1 vehicle injected at start of North bound through approach |
| **7** | 1 vehicle injected at start of West bound left approach |
| **8** | 1 vehicle injected at start of East bound left approach |
| **9** | 1 vehicle injected at start of South bound left approach |
| **10** | 1 vehicle injected at start of North bound left approach |

At each step of the simulation, the agent picks an action, $A_t$ from the action space and executes it. Initially, the agent performs random actions. However, as the agent proceeds with the training it learns the effect of performing a particular action given a particular environment state. Over time the agent is able to learn an optimal policy of vehicle injection.

**Reward function**

In order to ensure that the agent is learning effectively and reaching its goal, it is rewarded after executing each action. The aim of the attacker agent is to maximize the congestion at the subject intersection. Hence the reward function is based on the total waiting time for the current simulation step, and it is defined according to **equation 1**:

$$Reward_t = (gain * total\_waiting\_time_t) - (d * A_t) \tag{1}$$

A gain factor is multiplied with the total waiting time term in order to amplify or attenuate the impact of waiting time on the reward as required. An attacker will likely have budget constraints in executing the attack actions due to the cost of equipment required [6]. To account for the cost of injecting the fake vehicles, a penalizing term is deducted from the reward function. This term comprises of the current action executed, $A_t$ multiplied by a constant, $d$ to adjust the extent of penalty. Hence, higher order actions entail a higher penalty on the agent's reward which reflects the cost and difficulty of execution. In the simulation, a gain value of 1 is used and a value of 0.2 is used for the constant, $d$.

The agent seeks to learn a policy that helps it to maximize not only the reward received at current timestamp but also ensure that future cumulative rewards are maximized.

**DQN agent**

In reinforcement learning, the agent seeks to maximize not only its current reward but also the cumulative future rewards. It does so by following a policy, $\pi$ which it learns over the course of its training. Under this policy, the expected cumulative discounted returns corresponding to a state, $S_t$ and action, $A_t$ is given by a function called the action-value function [4] which is as follows:

$$q_\pi(s, a) = \mathbb{E}_\pi[G_t | S_t = s, A_t = a] = \mathbb{E}_\pi[\sum_{\kappa=0}^{\infty} \gamma^\kappa R_{t+\kappa+1} | S_t = s, A_t = a] \tag{2}$$

where, $\gamma$ is called the discount factor and $0 \leq \gamma \leq 1$. The discount factor is used to set the trade-off between immediate rewards and future rewards. If $\gamma = 0$, the agent chooses actions such that its immediate rewards are maximized without consideration for future rewards. On the other hand, setting $\gamma = 1$, makes the agent prioritize future rewards over current rewards and the actions are chosen accordingly.

Essentially, the agent is trained to learn or estimate this function by interacting with the environment. For simple environments and tasks, this can be accomplished by a q-table which maps each action and state pair to their corresponding expected future returns. However, for complex tasks requiring learning in an online setting, this approach becomes unfeasible. To resolve this, the use of deep neural networks was pioneered by the authors in [3] in which they taught an agent to play various video games. This approach is named the Deep Q-network (DQN) algorithm. Instead of mapping each action and state pair to determine the exact action-value function, the deep neural network is used to learn an approximation of the function. Such an agent is called a DQN agent. Since our problem involves interaction with a complex environment, we have adopted this agent in our attack modeling. **Table 2** summarizes the parameters of the agent used in our attack model:

**Table 2 Parameters of the RL agent**

| Parameter | Value |
|---|---|
| Number of layers | 4 |
| Number of hidden layers | 2 |
| Number of neurons in hidden layer 1 | 24 |
| Number of neurons in hidden layer 2 | 24 |
| Optimizer | ADAM |
| Loss function | Mean Absolute Error (MAE) |

## EXPERIMENTAL SET-UP

The implementation of the developed attack model was done in the Simulation of Urban Mobility (SUMO) traffic simulation package. SUMO not only enables the simulation of any network architecture but also provides functionalities to interact with the simulation via its Traffic Control Interface (TraCI) api. The TraCI api has a number of methods that can be used to retrieve values from the simulation as well as modify the elements of the simulation. In our attack simulation, TraCI was used to implement the control logic of the waiting time based adaptive traffic signal controller. Furthermore, the insertion of sybil vehicles was also done using TraCI's in-built functions for vehicle insertion.

### Simulation set-up

The roadway network used in our simulation is shown in **Figure 1**. The center intersection is our subject intersection, and it is where the attack is carried out. The traffic signal controller in the subject intersection is a waiting time based ATSC. In our network, each approach has three lanes: a dedicated left-turn lane, a dedicated through lane, and a shared through and right turn lane. The ATSC receives waiting time information from connected vehicles arriving at the subject intersection and switches phases accordingly. The subject intersection is connected to four adjacent intersections through the four different approach directions. Regular road vehicles that are part of the real-world traffic are originated from the approaches connected solely to these four adjacent intersections. A random demand schedule was prepared with pre-defined routes for the real-world traffic. Whilst the real-world traffic was defined prior to the simulation, the sybil vehicles were introduced into the simulation during run-time. The sybil vehicle insertion points were at the start of the approaches connected to the subject intersection as shown in **Figure 1** above.

### Waiting time based adaptive traffic signal controller

In order to demonstrate our attack, we have developed the waiting time-based adaptive signal control algorithm which allots green time if one of the eight traffic movements—east bound through (EBT), east bound left (EBL), west bound through (WBT), west bound left (WBL), south bound through (SBT), south bound left (SBL), north bound through (NBT), and north bound left (NBL)—experiences the longest average waiting time. Within the SUMO simulator, if a vehicle's speed is 0.1 m/s or less, it is deemed to have wait time. Once the vehicle starts to move and its speed exceeds 0.1 m/s, the waiting time is reset to zero. For each of the approaches of the subject intersection, the approach waiting time (AWT) was calculated using data collected via the in-built methods of the TraCI api. From this, the approach average waiting time (AAWT) of n vehicles is derived using the following equation:

$$AAWT_t = \frac{AWT}{n} \tag{3}$$

where n is the total number of vehicles heading toward the intersection in question, and $AAWT_t$ is defined as the average waiting time in seconds per vehicle at time step $t$. When computing the intersectional AAWT, the number n also indicates the total number of vehicles approaching the subject junction. A movement (through or left-turns) with the highest AAWT at time *t* relative to other movements is given green according to the waiting time-based ATSC. As a result, the green phase allocation is determined as follows:

$$\max(AAWT_t^M); where\ M = WBL, WBT, EBL, EBT, NBL, NBT, SBL, and\ SBT \qquad (4)$$

**Figure 2** depicts the flowchart for the waiting time based ATSC that is implemented in our simulation. Once the waiting time data are collected for all the approaches, the AAWT for each pair of conflicting movements are calculated. The movement pair having the highest AAWT is served the green phase. The phase allocation is reviewed every 5 seconds. If the age of the current green phase allocation is longer than 5 seconds, the ATSC recalculates the AAWT for conflicting movements. If a change is warranted, then the phase is updated. **Figure 3** presents waiting time per vehicle for the subject ATSC without any sybil attack. Note that average waiting time per vehicle is less than 10 seconds for all movements in each approach, which indicates that level of service for the subject intersection is A as per Highway Capacity Manual.

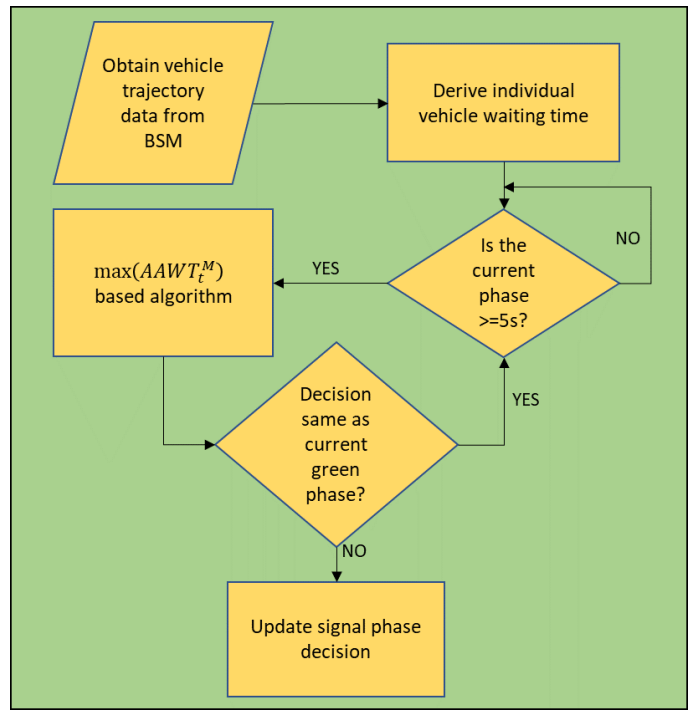

**Figure 2 Flowchart of waiting time based ATSC implementation**

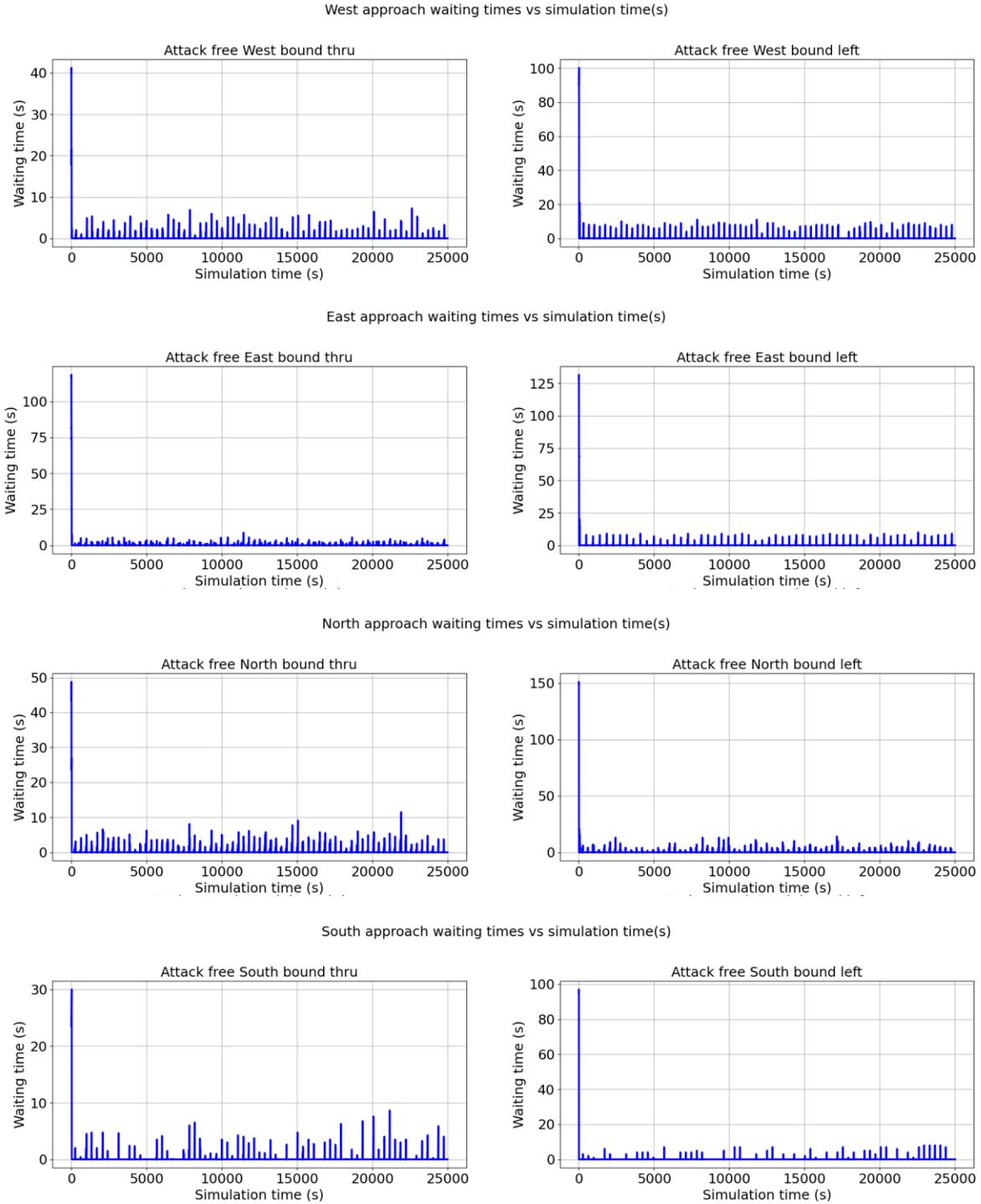

**Figure 3 Waiting time per vehicle for the subject ATSC without any sybil attack**

**Fake vehicle removal**

Once sybil vehicle are inserted, they follow the traffic rules just like the regular road vehicles. However, they have no physical existence but are merely nothing more than fake Basic Safety Messages (BSMs) that are broadcasted by the attacker RL agent to fool the ATSC. Hence, it is imperative that the sybil vehicles do not interfere with the flow of the regular traffic because in the real world, the physical vehicles will not see any sybil vehicles and will travel independently. While this condition is maintained during regular uncongested traffic flow, an issue arises whenever the sybil vehicles need to decelerate. Specifically, whenever the sybil vehicles in the simulation experience a red or yellow light, or there is a decelerating vehicle in front of it, the sybil vehicle is also decelerated by the simulator. This causes the real vehicles behind the sybil vehicles to decelerate as well, which is undesirable. Thus, in order to accurately model the real-world situation, it was necessary to resolve this.

The approach adopted to deal with this interference issue was to remove the fake vehicles whenever their deceleration values exceeded a certain threshold. If the sybil vehicles are removed as soon as they start to decelerate, then their time of contact with the ATSC would be insufficient to cause the attack. Furthermore, even during normal unobstructed travel when there are no red lights or vehicles in-front, all the vehicles experience slight deceleration and acceleration in order to maintain their speed in the way a human driver does. **Figure 4** shows the speed profile of a sample fake vehicle that was injected into the traffic stream. The region 1 represents the acceleration phase whereby the vehicle gains speed to reach the speed limit for that lane. After reaching the limit, the vehicles adjust their speed slightly above and below the speed limit to mimic how actual drivers maintain a constant speed. This is shown in region 2 of the speed profile. The last phase corresponds to deceleration in response to a red light ahead or a leader vehicle that is decelerating in-front of the ego vehicle.

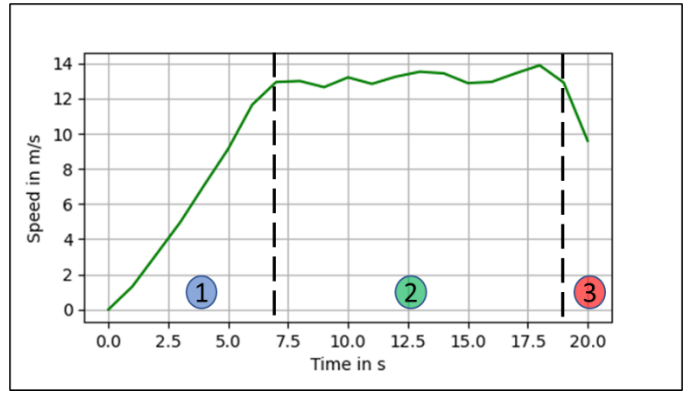

**Figure 4 Speed profile of a fake vehicle showing acceleration, constant speed and deceleration regions respectively**

If the removal condition is set as the start of deceleration, then the majority of the sybil vehicles get removed erroneously due to these slight perturbations in steady speed state. On the other hand, if the threshold is set too high then the interference with the physical road vehicles occurs. In particular, it was necessary to identify a removal threshold for the fake vehicles such that they are only removed if their speed profiles fall into region 3.

Accordingly, the acceleration and deceleration values were collected for all the sybil vehicle during one run of the simulation. **Figure 5(a)** shows the distribution of these values. From these values, we isolated only the values that corresponded to green light situation and no-vehicle in-front of the sybil vehicles. The distribution of the isolated acceleration values is shown in **Figure 5(b)**.

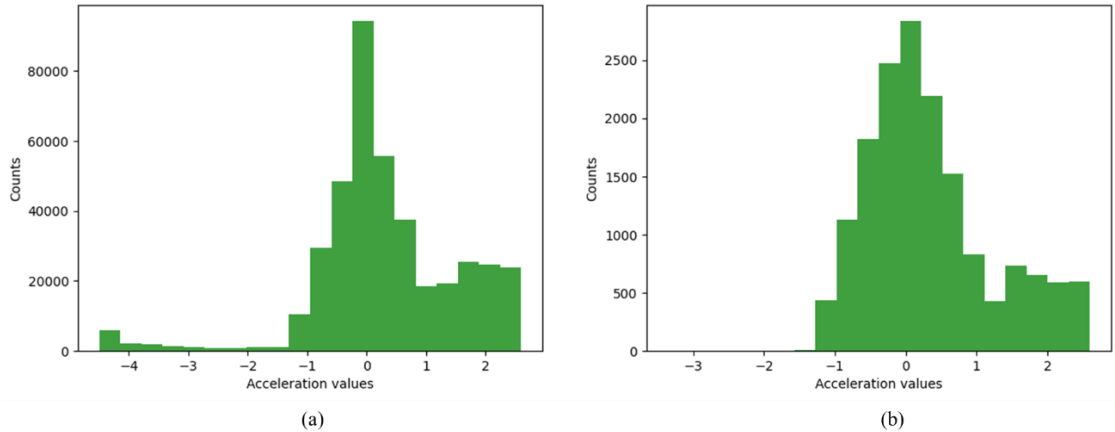

(a)                    (b)

**Figure 5 Distribution of acceleration values (a) Raw values and (b) Isolated values**

Next only the deceleration values were isolated. To ensure a fair trade off between attack impact and interference avoidance, the 5th percentile deceleration value was identified. This value was found to be 1.0174m/s$^2$. The interpretation of the value is that during normal unobstructed travel, 95% of the sybil vehicle experience deceleration whose magnitudes are less than 1.0174. This achieved a good compromise between the two extremes of the removal threshold.

**Baseline Scenario**

For the purpose of comparison between our model and the model developed in [9], a baseline simulation is set-up according to the model described in [9]. Both the simulations share the same network and demand schedules. However, the baseline simulation does not include vehicle injection in the left turn movement within its action space. Hence, the baseline agent only performs actions from 0 to 6 as shown in **Table 1**. The reward function for the baseline agent is also different from ours and is as follows:

$$Rewards_t = N_{halt} - N_{move} - (d * n)$$
(5)

Where $N_{halt}$ is the number of halted vehicles and $N_{move}$ is the number of moving vehicles, $n$ is the number of vehicles injection at timestep *t*, and *d* is a constant value. Since the baseline model does not account for the interference between real and fake vehicles, the fake vehicle removal is not implemented in the baseline simulation.

**RESULTS**

After setting-up the simulation, it is simulated and the RL agent is trained for a total of 100 episodes. Each training episode represents 1000 simulation steps of our traffic environment. At the end of the simulation, various metrics were collected and plotted to validate the developed attack model. At first a trial-and-error approach was used to identify the values of the hyperparameters to be used. Using this approach, a discount factor of 0.85 was determined for the future rewards. Additionally, a learning rate of 0.01 and replay memory of sample size 5000 were set. After determining the hyperparameters for our method, the results were compared against the baseline method after the same number of training episodes. The RL evaluation metrics have been split into two sets, one to validate that our RL attack agent is learning a policy with each training episode and another set to analyze the impact of the attack on the subject intersection.

**Agent Learning**

As the primary motivation of the agent is to maximize the congestion at the subject intersection, we recorded the total waiting time in each training episode. **Figure 6(a)** presents the plot of the total waiting time for each episode for our method. The figure shows that the total waiting time increases gradually over successive training episodes and reaches an almost steady value after 11 episodes. Hence, under the given conditions, the attacker RL agent is able to achieve a steady state congestion level within 11,000 simulation steps. Although the baseline method achieves higher total waiting times as shown in **Figure 6(b)**, it must be taken into account that the baseline model does not remove the fake vehicles once they are injected into the simulation. The baseline method also took around 11 episodes to reach a steady congestion level.

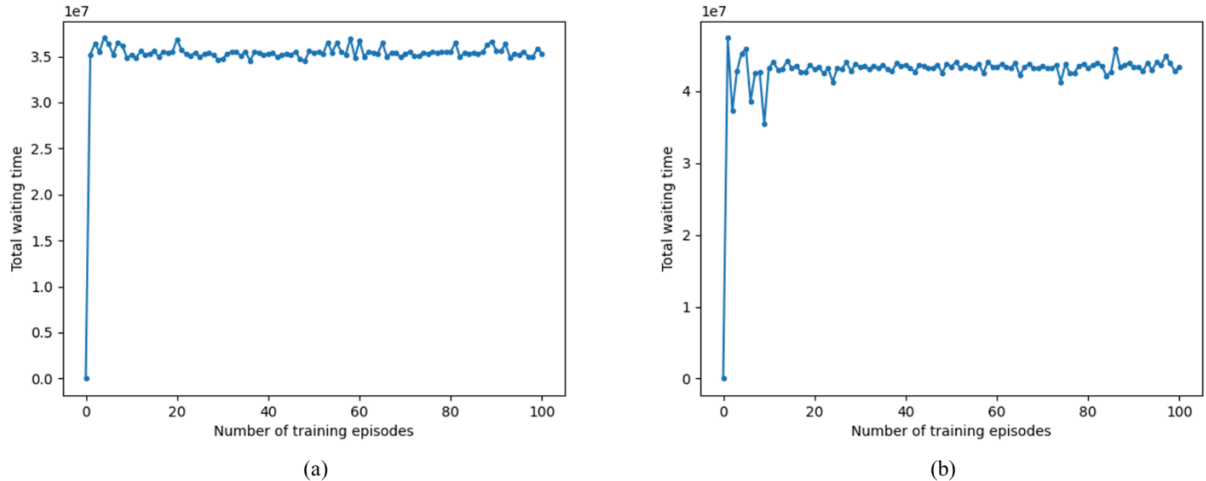

(a)                                          (b)

**Figure 6 Total waiting time at the subject intersection (a) Our model (b) Baseline model**

**Figure 7(a)** shows the total number of sybil vehicles that were injected into the traffic network in each episode. The graph has an initial spike at the end of the first episode after which, the number of sybil vehicles injected decreases and is maintained at a constant level of approximately 1000 vehicles. The initial spike may be attributed to the exploration stage of the agent. As there is a penalty term in the reward function that is proportional to the number of vehicles injected, injecting more vehicles than optimum will reduce the reward for the agent even if there is an increase in the waiting time. Hence, in subsequent training episodes, the agent injects a reduced number of vehicles. On the other hand for the baseline approach case, **Figure 7(b)** shows a wide fluctuation in the number of vehicles injected during the initial stages and stabilizes after around episode 30 at a higher vehicle count than our approach.

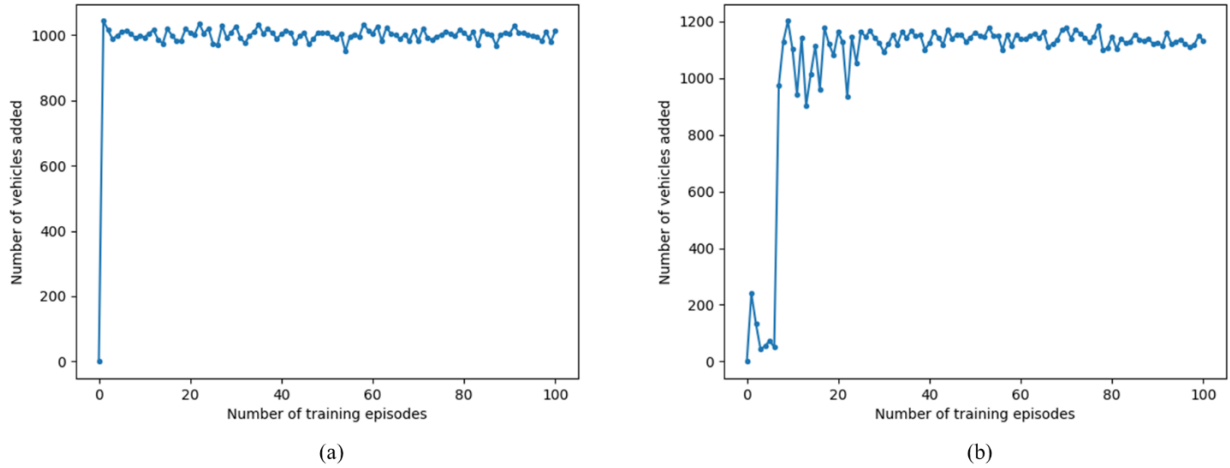

**Figure 7 Number of vehicles added into the subject intersection: (a) Our model (b) Baseline model**

The actions taken by the agent in each simulation step is plotted against the corresponding rewards obtained in **Figure 8(a)**. It shows that the agent takes action 4 most frequently compared to other actions. In contrast, actions 3, 5 and 7 are not taken at all. However, **Figure 8(b)** shows that the baseline agent takes all six actions evenly throughout its action space with action 2 being the most frequently taken.

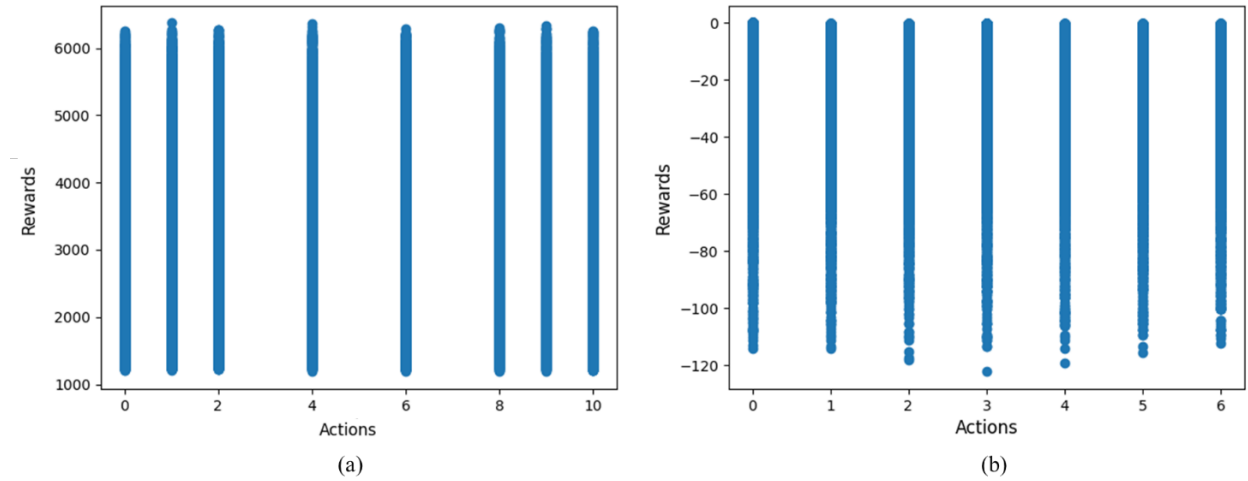

**Figure 8 Agent actions performed against the obtained rewards (a) Our model (b) Baseline model**

**Attack Impact**

In addition to the agent learning performance, the impact of the attack needs to be assessed. **Figures 9** and **10** present the waiting time for the last episode of training which consists of data for 1000 simulation steps for all four approaches. The impact of the attack on the waiting time can only be understood in comparison with the attack free scenario. Hence, the plots also include the attack free scenario on top of the under-attack scenario. In addition, the results for the baseline approach are presented as well for comparison against our method. The results show that the agent is able to cause a drastic increase in the waiting time at the different approaches and cause a deterioration in the performance.

**Figure 9** presents the waiting time plot for the through movements for each of the four approaches. Compared to the attack free scenario, both our method and the baseline method yielded significant congestion in terms of waiting time increase. However, the baseline method had a better performance in terms of waiting time increase. Especially, in the west bound through approach, the baseline method was noticeably more effective.

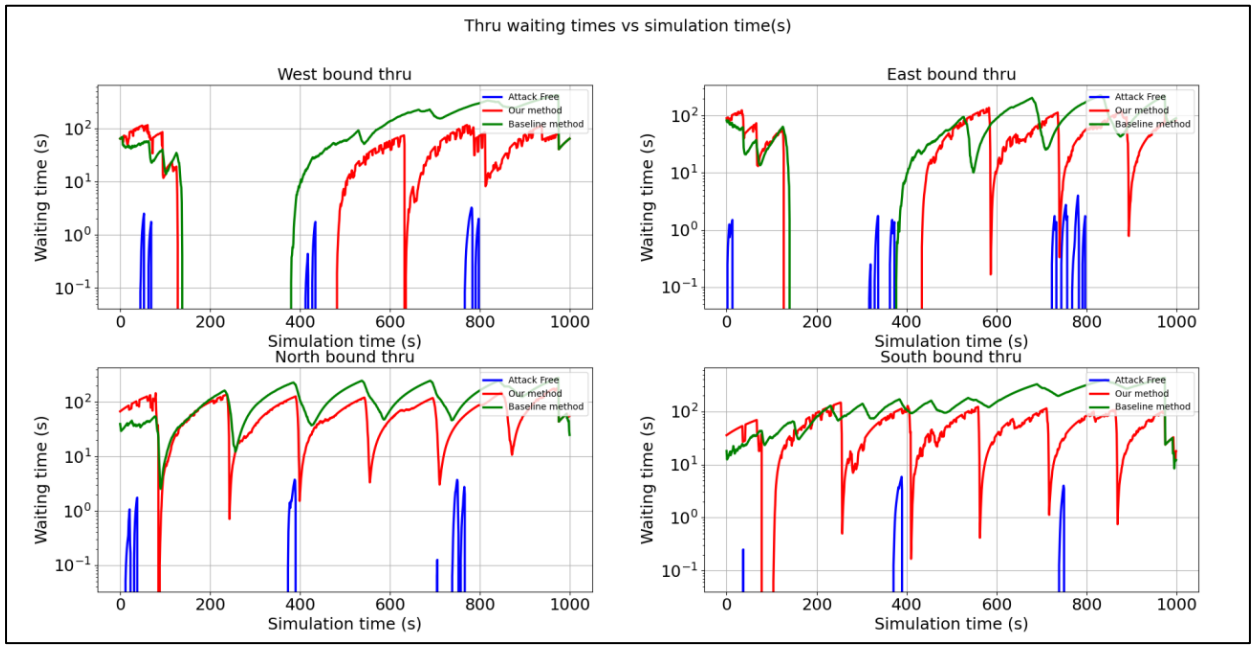

**Figure 9 Waiting time plots for through movements**

The waiting time plots for the left turn movement are presented in **Figure 10**. In this instance, our method performs equally well compared to the baseline. Notably, the peak waiting times are the same in both methods. Furthermore, in the south bound approach, our method was able to create longer periods of congestion. In consolidation, while the baseline method was more effective in creating congestion in the through movements, our developed attack model was able to generate equal amount congestion in the left turn movements. Thus, the expansion of the action space in our attack model to include vehicle injections in the left turn movements was validated. Furthermore, although the baseline method has consistently higher waiting times in the through movements, it does not account for the interference issue in the simulation between the fake vehicles and regular traffic flow.

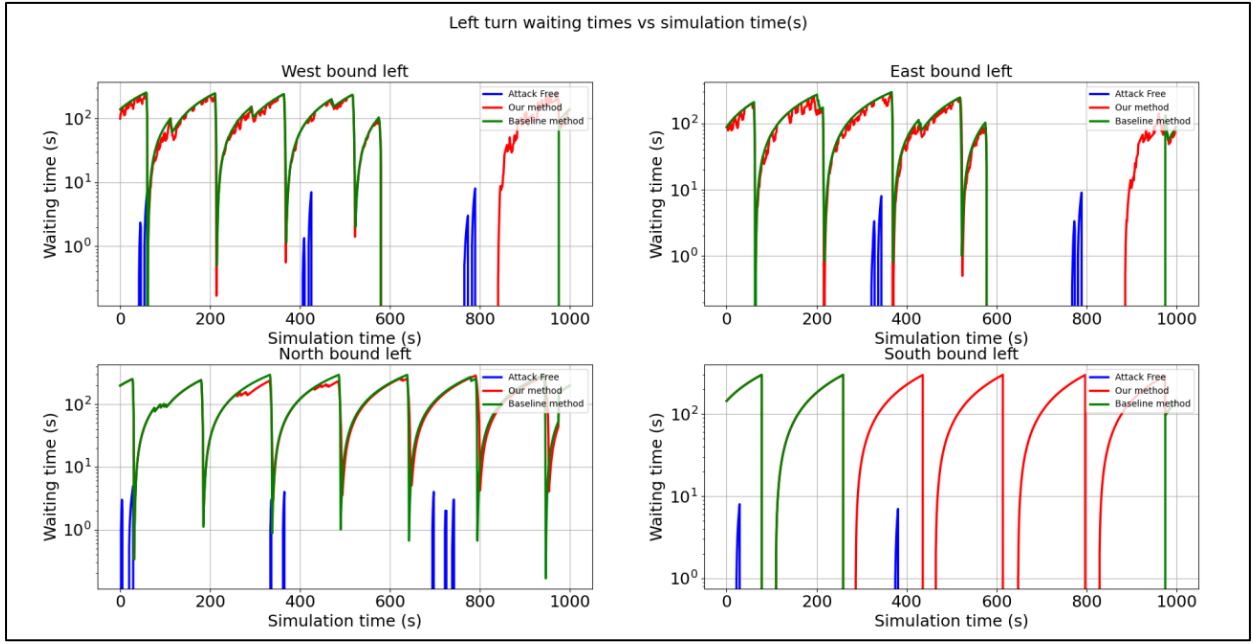

**Figure 10 Waiting time plots for left turn movements**

Similarly, the accumulated waiting time plots for the different approaches are presented in **Figures 11** and **12**. In SUMO, the accumulated waiting time of a vehicle measures the sum of the waiting time of that vehicle over the last 100 seconds. Therefore, unlike the waiting time which by definition gets reset to zero as soon as the vehicle starts to move, the accumulated waiting time for a lane does not get reset to zero. The accumulated waiting time is able to capture the total time that a vehicle incurred waiting time as it moves along a particular lane.

The accumulated waiting time plots for the through movements are presented in **Figure 11**. It follows the same trend as the waiting time plots with the baseline method performing the best. The biggest difference is observed in the south bound through approach. However, although the baseline method performs better, it must be noted that the baseline method does not implement a removal strategy for the inserted fake vehicles. This would cause the real vehicles to be congested behind fake vehicles thereby accumulating waiting time, thus yielding the higher values.

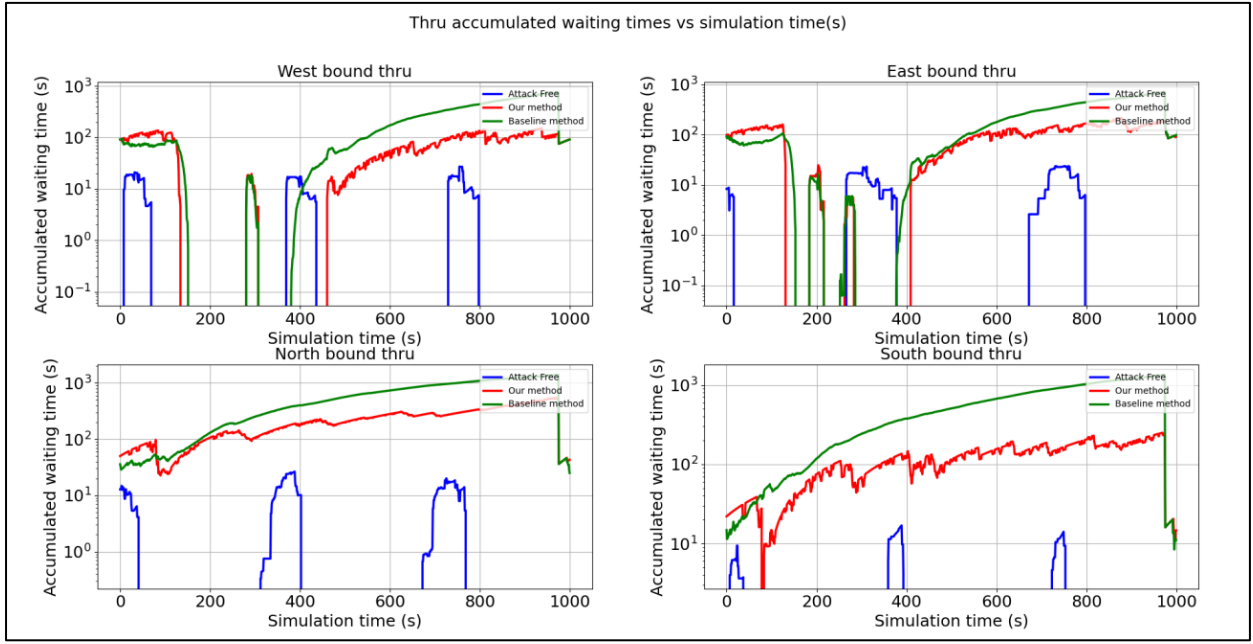

**Figure 11 Accumulated waiting time plots for through movements**

However, for the left turn movements, our method performed comparably with the baseline method. In converse with the through movements case, our method performed better in the south bound left turn approach as the vehicles accumulated waiting time for greater number of periods as shown in **Figure 12**.

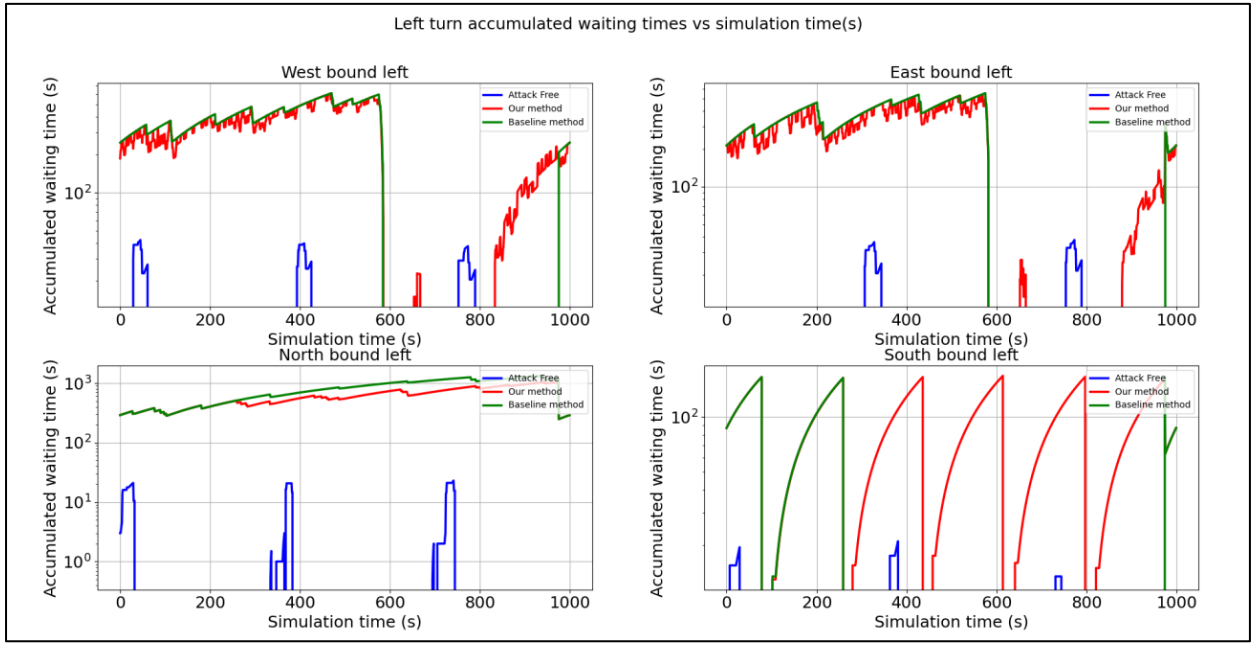

**Figure 12 Accumulated waiting time plots for left turn movements**

The results prove that our developed attack model is effective in creating congestion in all the approaches. In comparison with the baseline method, it would seem that it performs better than our developed method. However, by targeting the left turn movements and the implementing a fake vehicle removal strategy, our model has been made more realistic and feasible for deployment. This increases the difficulty of detection and thus makes this attack superior compared to the baseline method. Furthermore, it is demonstrated that even without violating traffic rules it is possible for an attacker to achieve their objective.

## CONCLUSION AND FUTURE WORK

This study demonstrates a novel attack model against waiting time based ATSC systems by leveraging RL techniques. We have developed a simulation environment within which the attacking RL agent is trained. The agent attacks the ATSC by inserting fake vehicles at the various approaches connected to the subject intersection in order to fool the ATSC and cause it to produce suboptimal signal timing and hence introduce congestion into the system. Our analyses revealed that the RL agent can learn an optimal policy for creating an intelligent attack, and our attack model was successfully able to achieve the objective of congestion creation at the subject intersection. Further works on this model can help make it more robust and universal. Firstly, the model can be tested on different intersection architectures having variable number of lanes and movement directions. Secondly, the developed model should also be deployed against other types of ATSCs, such as DRL based ATSCs. The learning from these different cases can be used to develop an attack framework for testing the vulnerabilities of any existing or emerging ATSCs. The next step for this research is to develop an attack detection and mitigation framework to defend against such intelligent automated attacks on ATSCs. It is anticipated that the current work will help to spur the research in that direction and make ATSCs more secure and help to increase its widespread deployment.